\documentclass[aps,showpacs,prl,twocolumn,footinbib,superscriptaddress]{revtex4-1}
\usepackage{epsfig}
\usepackage{amsfonts}
\usepackage{amsmath}
\usepackage{bm}
\usepackage{xcolor}
\usepackage{newtxtext,newtxmath}
\usepackage{booktabs}
\usepackage{yfonts}

\usepackage[utf8]{inputenc}
\usepackage[colorlinks,bookmarks=false,citecolor=blue,linkcolor=blue,urlcolor=blue]{hyperref}
\usepackage{soul}
\usepackage{stackengine} 

\usepackage{changepage}

\newcommand{\bra}[1]{\left< #1 \right|}
\newcommand{\ket}[1]{\left| #1 \right>}
 
\newcommand{\ketbra}[2]{\left| #1 \right> \left< #2 \right|}
\newcommand{\av}[1]{\left< #1 \right>}

\newcommand{\upstate}{\ket{\uparrow}}
\newcommand{\downstate}{\ket{\downarrow}}

\newcommand{\be}{\begin{equation}}
\newcommand{\ee}{\end{equation}}

\begin{document}

\title{Engineering non-binary Rydberg interactions via phonons in an optical lattice}
\author{F. M. Gambetta}
\affiliation{School of Physics and Astronomy, University of Nottingham, Nottingham, NG7 2RD, United Kingdom}
\affiliation{Centre for the Mathematics and Theoretical Physics of Quantum Non-equilibrium Systems, University of Nottingham, Nottingham NG7 2RD,  United Kingdom}
\author{W. Li}
\affiliation{School of Physics and Astronomy, University of Nottingham, Nottingham, NG7 2RD, United Kingdom}
\affiliation{Centre for the Mathematics and Theoretical Physics of Quantum Non-equilibrium Systems, University of Nottingham, Nottingham NG7 2RD,  United Kingdom}
\author{F. Schmidt-Kaler}
\affiliation{QUANTUM, Institut f\"ur Physik, Johannes Gutenberg-Universit\"at Mainz, 55128 Mainz, Germany}
\affiliation{Helmholtz-Institut Mainz, 55128 Mainz, Germany}
\author{I. Lesanovsky}
\affiliation{School of Physics and Astronomy, University of Nottingham, Nottingham, NG7 2RD, United Kingdom}
\affiliation{Centre for the Mathematics and Theoretical Physics of Quantum Non-equilibrium Systems, University of Nottingham, Nottingham NG7 2RD,  United Kingdom}

\date{\today}

\begin{abstract}
	Coupling electronic and vibrational degrees of freedom of Rydberg atoms held in optical tweezer arrays offers a flexible mechanism for creating and controlling atom-atom interactions. We find that the state-dependent coupling between Rydberg atoms and local oscillator modes gives rise to two- and three-body interactions which are controllable through the strength of the local confinement. This approach even permits the cancellation of two-body terms such that three-body interactions become dominant. We analyze the structure of these interactions on two-dimensional bipartite lattice geometries and explore the impact of three-body interactions on system ground state on a square lattice. Focusing specifically on a system of $ ^{87} $Rb atoms, we show that the effects of the multi-body interactions can be maximized via a tailored dressed potential within a trapping frequency range of the order of a few hundred kHz and for temperatures corresponding to a $ >90\% $ occupation of the atomic vibrational ground state. These parameters, as well as the multi-body induced time scales, are compatible with state-of-the-art arrays of optical tweezers.  Our work shows a highly versatile handle for engineering multi-body interactions of quantum many-body systems in most recent manifestations on Rydberg lattice quantum simulators.
\end{abstract}

\maketitle

\textit{\textbf{Introduction.}---} 
In the past years Rydberg atoms~\cite{Gallagher:2005,Saffman:2010,Low:2012} held in optical tweezer arrays have emerged as a new platform for the implementation of quantum simulators and, potentially, also quantum computers~\cite{ Saffman:2016,Gross:2017,Bernien:2017,Browaeys:2016,Jau:2015,Lienhard:2018,Omran:2019}. One-~\cite{Bernien:2017}, two-~\cite{deMello:2019} and three-dimensional~\cite{Barredo:2018} arrays containing hundreds of qubits are in principle achievable and the wide tunability of Rydberg atoms grants high flexibility for the implementation of a whole host of quantum many-body spin models. The physical dynamics of these quantum simulators takes place in the electronic degrees of freedom which mimic a (fictitious) spin particle. Effective magnetic fields and interactions are achieved via light-shifts effectuated by external laser fields and the electrostatic dipolar interaction between Rydberg states. Additional tuning with electric~\cite{Vogt:2007} and magnetic fields~\cite{Pohl:2009} permits the realization of exotic interactions, allowing for the study of ring-exchange Hamiltonians~\cite{Glaetzle:2014,vanBijnen:2015,Glaetzle:2017,Weimer:2010}, frustrated-spin models~\cite{Balents:2010,Glaetzle:2015,Whitlock:2017} or crystallization phenomena~\cite{Pohl:2010,Schachenmayer:2010,Schauss:2015}. Within this context, in the last decade systems with tunable two- and three-body interactions~\cite{Pachos:2004,Buchler:2007,Schmidt:2008,Capogrosso:2009,Bonnes:2010} have attracted a lot of attention since the latter are responsible for the emergence of many exotic quantum states of matter, ranging from topological phases~\cite{Moore:1991,Kitaev:2006} to spin liquids~\cite{Moessner:2001,Misguich:2002}.

In this work we put forward a new mechanism for engineering non-binary interactions in Rydberg tweezer arrays~\cite{Kaufman:2012,Nogrette:2014,Labuhn:2014, Barredo:2016,Bernien:2017,Lienhard:2018,Labuhn:2016,Endres:2016,Marcuzzi:2017,Cooper:2018,Levine:2018, Saskin:2019}.
Here, each atom is held in place by a strong local harmonic potential. The simultaneous excitation of neighboring atoms to the Rydberg state gives rise to a mechanical force that couples the electronic degrees of freedom to the local phonon modes. We show that this coupling gives rise to effective spin-spin interactions between excited atoms. Similar mechanisms in which effective inter-particle interactions arise as a consequence of the coupling with an extra degree of freedom have been extensively studied in condensed matter systems. Here, well-known examples include the electron-electron interaction mediated by lattice phonons in metals~\cite{Mahan:2000} and the indirect spin-spin couplings~\cite{Yosida:1996,*Nolting:2009} due to the Ruderman-Kittel-Kasuya–Yosida~\cite{Ruderman:1954}, superexchange~\cite{Anderson:1950}, and Dzyaloshinskii–Moriya mechanisms~\cite{Dzyaloshinsky:1958,*Moriya:1960}. In these cases, integrating out the extra degree of freedom typically results in two-body effective interactions between the remaining degrees of freedom. Crucially, in our system, since spins and phonons are coupled via pairs of Rydberg atoms, not only two-body but also three-body effective interactions arise. We analyze in depth the interplay between the various effective couplings in the case of two-dimensional (2D) bipartite lattice geometries, demonstrating that regimes dominated by three-body interactions can be achieved. Our results show that the multi-body interactions arising from the electron-phonon coupling are highly tunable and can drive non-trivial phase transitions in the ground state of a Rydberg spin system. By tuning the local harmonic potentials, we show that checkerboard, striped, and clustered phases occur as well as signatures of frustration phenomena. Our work is directly relevant for recent developments on the domain of quantum simulation with Rydberg tweezer arrays where it highlights a so far unanticipated mechanism for experimentally realizing exotic interactions.

\textit{\textbf{2D model.}---} 
\begin{figure}[t]
	\centering
	\includegraphics[width=\columnwidth]{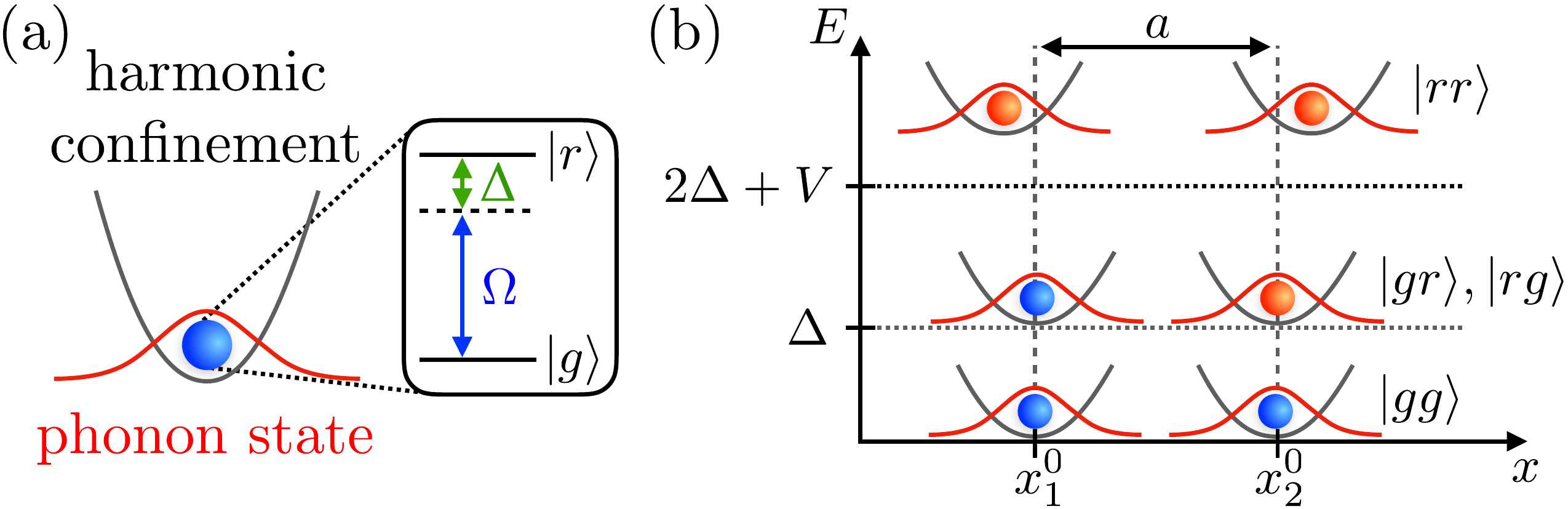}
	\caption{{\bf Setup}. (a) Each atom is modeled as a two-level system with ground state $ |g\rangle $ and excited Rydberg state $ |r\rangle $. The two levels are coupled by a laser with Rabi frequency $ \Omega $ and detuning $ \Delta $. The atom is trapped inside a tight harmonic optical tweezer (grey) and, at low temperature, it occupies the ground state of the associated phonon degree of freedom (red). For simplicity, we assume that Rydberg and ground state experience the same trapping potential. (b) Energy diagram of a two atom system arranged along the $ x- $axis. When both atoms are excited to the Rydberg state, $ |r r \rangle $, they experience, in addition to the electronic dipolar interaction $ V $, the potential change $ \delta V $ arising as a consequence of the coupling between spin and phonon degrees of freedom and consisting of both two- and three-body contributions; see text for details. This also results in a state-dependent displacement $ \delta x_{1,2} $ of the atoms from their equilibrium position $ x^0_{1,2} $, separated by the lattice spacing $ a$.  }
	\label{fig:setup}
\end{figure}
We consider a 2D lattice of $ N $ Rydberg atoms in the $ x-y $ plane, whose sites are labeled by $ \bm{k}=(k_x,k_y) $. The electronic degree of freedom is modeled as effective two-level system (with $\downstate $ and $ \upstate $ denoting the ground state and the Rydberg excited state, respectively)~\cite{Low:2012,Marcuzzi:2017}. The two levels are coupled by a laser with Rabi frequency $ \Omega $ and detuning $ \Delta $ [see Fig.~\ref{fig:setup}(a)].
Each of the atoms, with mass $ m $, is trapped in a strong three-dimensional harmonic potential, characterized by trapping frequencies $ \omega_\mu $ along the directions $ \mu=x,y,z $. The atomic motion inside the confining potential can then be described in terms of the bosonic operators $ b_{\bm{k},\mu} $. The Hamiltonian describing the single-particle dynamics is
\begin{equation}\label{eq:Hterms}
	H_\mathrm{sp}=\sum_{\mu=x,y,z}\sum_{\bm{k}} \hbar \omega_\mu b^{\dagger}_{\bm{k};\mu} b_{\bm{k};\mu}+\sum_{\bm{k}}\left[\Omega\sigma^x_{\bm{k}}+\Delta n_{\bm{k}}\right].
\end{equation}
Here, $ n_{\bm{k}}=(1+\sigma^z_{\bm{k}})/2 $ and $ \sigma^{\mu}_{\bm{k}} $ are the Rydberg number operator  and Pauli matrices acting on the atom at site $ \bm{k} $ and position $ \bm{r}_{k}=(x_k,y_k,z_k) $, respectively. Any two atoms at lattice positions $ \bm{r}_{\bm{k}} $ and $ \bm{r}_{\bm{m}} $, if excited to the Rydberg state, interact through the two-body potential $ V(\bm{r}_{\bm{k}},\bm{r}_{\bm{m}}) $, which depends on the inter-particle distance $ |\bm{r}_{\bm{k}}-\bm{r}_{\bm{m}}| $~\cite{Gallagher:2005,Low:2012,Browaeys:2016}. The overall Hamiltonian is therefore $ H=H_\mathrm{sp}+H_\mathrm{int} $ with 
\begin{equation}
H_\mathrm{int}=\sideset{}{'}\sum_{\bm{k},\bm{m}} V(\bm{r}_{\bm{k}},\bm{r}_{\bm{m}})n_{\bm{k}} n_{\bm{m}}, \label{eq:Hint}
\end{equation}
where the prime in the sum implies that terms with equal indices are excluded. Note that in Eq.~\eqref{eq:Hterms} we have assumed  the same trapping frequencies $ \omega_ \mu $ for atoms in the ground and in the Rydberg state. This “magic” condition can be realized in Rydberg tweezer arrays through bottle beam traps~\cite{Barredo:2019}. Furthermore, a small frequency mismatch between the two states does not affect our central results, as discussed in the Supplemental Material (SM)~\cite{Note1}.

At low temperature each atom oscillates around the minimum of its local potential, $ \bm{r}^0_{\bm{k}} $, and its position can thus be written as $ \bm{r}_{\bm{k}}=\bm{r}^0_{\bm{k}}+\delta \bm{r}_{\bm{k}} $,  with $ \delta \bm{r}_{\bm{k};\mu}=\ell_{\mu}(b^\dagger_{\bm{k}; \mu}+b_{\bm{k}; \mu}) $ being the atomic displacements from equilibrium. Here, $ \bm{\ell}=(\ell_x,\ell_y,\ell_z) $ is the vector of the characteristic lengths associated with the harmonic trapping potentials in the three spatial directions with $ \ell_{\mu} = \sqrt{\hbar(2m\omega_\mu)^{-1}} $. As a consequence, the two-body interaction depends on the displacements: $ V(\bm{r}^0_{\bm{k}}+\delta\bm{r}_{\bm{k}},\bm{r}^0_{\bm{m}}+\delta\bm{r}_{\bm{m}}) n_{\bm{k}} n_{\bm{m}} $. Clearly, this implies that a coupling between electronic and vibrational degrees of freedom emerges. 

This situation is reminiscent of a mechanism for creating long-range spin models in arrays of trapped ions~\cite{Porras:2004,Blatt:2012, Vogel:2019}. In that case, the interplay of long-range Coulomb repulsion between the ions and laser induced spin-dependent forces results in an effective long-range spin-spin interaction and allows to simulate a rich variety of quantum systems. However, in contrast to the ions, Eq.~\eqref{eq:Hint} implies that in our setup the potential $ V(\bm{r}_{\bm{k}},\bm{r}_{\bm{m}}) $ couples electronic and vibrational degrees of freedom only when two atoms are excited, which is the origin of many-body spin interaction terms.  

To demonstrate this, we focus on the strong confinement regime, in which the displacements $\delta \bm{r}_{\bm{k}}$ are much smaller than inter-atomic distances. Indeed, this represents the typical situation in Rydberg quantum simulators~\cite{Bernien:2017,Labuhn:2016,Lienhard:2018,Endres:2016}. By expanding the potential in Eq.~\eqref{eq:Hint} in a Taylor series to the first order in $ \delta \bm{r} $, the atom-atom interaction Hamiltonian acquires the form
\begin{equation}\label{eq:Hint_W}
H_{\mathrm{int}}=\sideset{}{'}\sum_{\bm{k},\bm{m}} \left[V^0_{\bm{k},\bm{m}}+\sum_{\mu}W_{\bm{k},\bm{m};\mu}\left(b^\dagger_{\bm{k}; \mu}+b_{\bm{k}; \mu}\right)\right] n_{\bm{k}} n_{\bm{m}},
\end{equation}
where $  V^0_{\bm{k},\bm{m}}\equiv V(\bm{r}^0_{\bm{k}},\bm{r}^0_{\bm{m}}) $ and 
\begin{equation}\label{eq:W}
W_{\bm{k},\bm{m};\mu} = 2 \ell_{\mu} \left[\nabla_{\bm{r}_{\bm{k}}} V(\bm{r}_{\bm{k}},\bm{r}^0_{\bm{m}})|_{\bm{r}_{\bm{k}}=\bm{r}^0_{\bm{k}}}\right]_\mu
\end{equation}
Finally, since the spin-phonon coupling in Eq.~\eqref{eq:Hint_W} is linear in the bosonic operators, we can apply a polaron transformation, $ U $ (see SM~\cite{Note1}), to decouple spin and phonon dynamics. We obtain~\cite{Porras:2004,Deng:2005,Note1}
\begin{equation}\label{eq:H_CS}
U H U^\dagger = H_\mathrm{sp}+H_\mathrm{2B}+H_\mathrm{3B}+H_\mathrm{res}+O(\ell_{\mu}^2/a^2),
\end{equation}
with 
\begin{subequations}\label{eq:Hterms_CS}
	\begin{align}
    H_\mathrm{2B}&=\sideset{}{'}\sum_{\bm{k},\bm{m}} \left(V^0_{\bm{k},\bm{m}}-\widetilde{V}_{\bm{k};\bm{m}}\right)n_{\bm{k}} n_{\bm{m}},\label{eq:H2B}\\
	H_\mathrm{3B}&=-\sideset{}{'}\sum_{\bm{k},\bm{p},\bm{q}} \widetilde{V}_{\bm{k};\bm{p},\bm{q}} n_{\bm{k}} n_{\bm{p}} n_{\bm{q}}\label{eq:H3B}.
	\end{align}
\end{subequations}
Here, we have introduced the coefficients $ \widetilde{V}_{\bm{k};\bm{p},\bm{q}}=\sum_{\mu} (\hbar \omega_\mu)^{-1} W_{\bm{k},\bm{p}; \mu} W_{\bm{k},\bm{q}; \mu}  $ and $ \widetilde{V}_{\bm{k};\bm{m}}\equiv\widetilde{V}_{\bm{k};\bm{m},\bm{m}} $. Equations~\eqref{eq:H2B} and \eqref{eq:H3B} show that, as consequence of the spin-phonon coupling, an effective atom-atom interaction emerges. The latter consists of an extra two-body [Eq.~\eqref{eq:H2B}] and a novel three-body term [Eq.~\eqref{eq:H3B}], whose strengths are both $ \propto \widetilde{V}_{\bm{k};\bm{p},\bm{q}} $. Importantly, the coefficients $ \widetilde{V}_{\bm{k};\bm{p},\bm{q}}  $ depend on the trapping frequencies $ \omega_\mu $ and are therefore tunable via the harmonic confinement.

The term $ H_\mathrm{res} $ in Eq.~\eqref{eq:H_CS} describes a residual spin-phonon coupling, which is negligible in the limit $ |W_{\bm{k},\bm{m};\mu}| \ll\hbar\omega_\mu $ ~\cite{Porras:2004,Deng:2005,Blatt:2012, Note1}. In this regime the phonon dynamics decouples from the spins. The approximation further improves at temperatures low enough to ensure a $ \gtrsim 90\% $ population in the vibrational ground state.  Such temperatures can be experimentally achieved in state-of-the-art optical tweezers via Raman sideband cooling~\cite{Kaufman:2012,Thompson:2013}. Details on the validity of this spin-phonon decoupling approximation are provided in next section and in SM~\cite{Note1}.

\begin{figure}[t]
	\centering
	\includegraphics[width=\columnwidth]{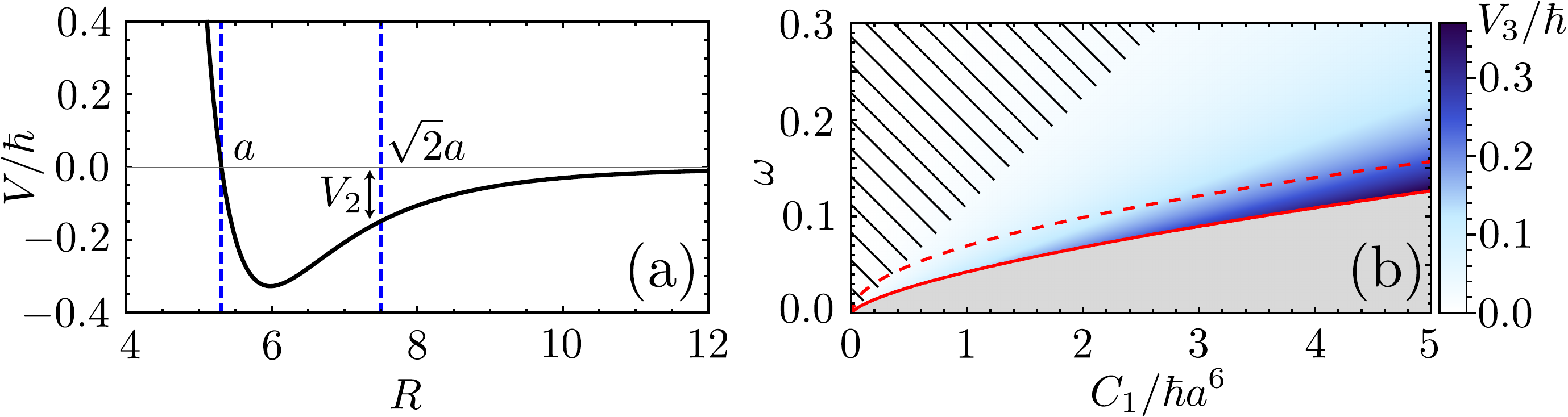}
	\caption{{\bf MW-dressed potential and three-body interaction strength.} (a) Dressed potential $ V/\hbar $ (units $ 2\pi\ \times $ MHz) as a function of inter-atom distance $ R $ (units $ \mu $m) obtained via MW dressing. See text and SM~\cite{Note1}. In the square lattice case, the atomic separations are $a\approx 5.3\mu \mathrm{m}$ and $ a_{\mathrm{NNN}}=\sqrt{2}a $ (blue dashed). The functional form of the potential is described by Eq.~\eqref{eq:compositeV}, with $ C_1/(\hbar a^6)=2\pi\times2.6 $ MHz, $ C_2/(\hbar a^6)=2\pi\times0.3$ MHz, $ c_1\approx-C_1 $, and $ c_2=0 $. (b) Density plot of $ V_3/\hbar $ as a function of $ C_1/(\hbar a^6)$ and $ \omega $ (units $ 2\pi\ \times $ MHz), with $ \omega_\mu=\omega $. The regime with $ |W_{\bm{k},\bm{m}; \mu }| \leq \hbar \omega_\mu $ (in gray) is separated by the bound given in Eq.~\eqref{eq:approx_omega} (red solid curve). The case with $ V_3=V_2 $ is indicated by the red dashed curve. The hatched area denotes the regime where the time scale corresponding to $ V_3 $ is $ >50\ \mu $s. Here, $ C_2/(\hbar a_\mathrm{NNN}^6)=-0.1 C_1/(\hbar a^6) $, $ c_1=-C_1 $, and $ c_2=0 $.} 
	\label{fig:approx}
\end{figure}

\textit{\textbf{Microwave dressed Rydberg states.}---}
The strength of the phonon-mediated effective interactions in Eqs.~\eqref{eq:H2B} and \eqref{eq:H3B} is directly connected to the strength of the dipolar ones: This is because the coefficients $ W_{\bm{k},\bm{m};\mu} $ are proportional to the gradient of $ V(\bm{r}_{\bm{k}},\bm{r}_{\bm{m}}) $. Typical dipolar interactions exhibit a power-law behavior $ \propto |\bm{r}_{\bm{k}}-\bm{r}_{\bm{m}}|^{-\alpha}$ (e.g., $ \alpha=6 $ for a van der Waals potential). In which case, one generally finds that $ V(\bm{r}_{\bm{k}},\bm{r}_{\bm{m}}) \gg \widetilde{V}_{\bm{k};\bm{m}} $. This means that, in common situations, phonon-mediated interactions only represent a small correction. However, the interaction potential between excited atoms can be tailored via microwave (MW) dressing of two different Rydberg states~\cite{Sevincli:2014,Marcuzzi:2015,Note1}, allowing to make the effective interactions dominant. In Fig.~\ref{fig:approx}(a) we show one possible realization of such potential, obtained via MW dressing of the atomic levels $|65S\rangle$ and $ |75P\rangle$ of $ {}^{87} $Rb atoms arranged on a square lattice. Here, $ a $ and $ a_{\mathrm{NNN}}=\sqrt{2} $ are the distances at equilibrium between nearest neighbors (NNs) and next-nearest neighbors (NNNs), respectively. By properly choosing the MW field parameters (see SM for details~\cite{Note1}), the potential can be parameterized, to a good degree of approximation, as
\begin{equation}\label{eq:compositeV}
V(\bm{r}_{\bm{k}}, \bm{r}_{\bm{m}})\approx\begin{cases}
\frac{C_{1}}{2|\bm{r}_{\bm{k}}-\bm{r}_{\bm{m}}|^6} + \frac{c_{1}}{2 a^6}& \text{for }|\bm{r}_{\bm{k}}-\bm{r}_{\bm{m}}|\approx a,\\
\frac{C_{2}}{2|\bm{r}_{\bm{k}}-\bm{r}_{\bm{m}}|^6} + \frac{c_2}{2 (a_{\mathrm{NNN}})^6} & \text{for }|\bm{r}_{\bm{k}}-\bm{r}_{\bm{m}}|\approx a_{\mathrm{NNN}},
\end{cases}
\end{equation}
with, for a typical dressed potential, $ V(\bm{r}_{\bm{k}},\bm{r}_{\bm{m}})\approx0 $ for $ |\bm{r}_{\bm{k}}-\bm{r}_{\bm{m}}|>a_{\mathrm{NNN}} $~\cite{Note1}. 
MW dressing allows to control the values of the constants $ C_{1,2} $ and $ c_{1,2} $ in Eq.~\eqref{eq:compositeV} independently and, in turn, to tune the strength of the dipolar potential (as well as its gradient) at NN and NNN distances, denoted by $ V_1 $ and $ V_2 $, respectively.

\textit{\textbf{Phonon-mediated interactions.}---} For the case shown in Fig.~\ref{fig:approx}(a), we have $ V_1/\hbar\approx 0 $, $ V_2/\hbar \approx 2\pi\times 0.3 $ MHz, and $ \hbar^{-1}dV/dR|_{R=a}=2\pi\times1.45 $ MHz. In this way we can thus achieve regimes dominated by the phonon-mediated interactions, whose strength along the $ \mu $ direction is described by the parameter
\begin{equation}\label{eq:V3}
V_{3,\mu}=\frac{36 \ell_{\mu}^2}{\hbar \omega_\mu a^2} \left(\frac{C_1}{a^6}\right)^2.
\end{equation}
In this case, Eqs.~\eqref{eq:H2B} and \eqref{eq:H3B} become
\begin{subequations}\label{eq:Hterms_NNN}
	\begin{align}
	H_\mathrm{2B}&=\sum_{\langle \bm{k},\bm{m}\rangle} \left(V_1  -  \widetilde{V}_{\bm{k};\bm{m}} \right)n_{\bm{k}} n_{\bm{m}}+\sum_{\langle\langle \bm{k},\bm{m} \rangle\rangle} V_2 n_{\bm{k}} n_{\bm{m}},\label{eq:H2BNNN}\\
	H_\mathrm{3B}&=-\sum_{\langle\bm{k},\bm{p},\bm{q}\rangle} \widetilde{V}_{\bm{k};\bm{p},\bm{q}} n_{\bm{k}} n_{\bm{p}} n_{\bm{q}}\label{eq:H3BNNN},	\end{align}
\end{subequations}
where, explicitly, $ \widetilde{V}_{\bm{k};\bm{p},\bm{q}}=  V_{3,\mu} \tilde{R}^0_{\bm{k},\bm{p}; \mu} \tilde{R}^0_{\bm{k},\bm{q}; \mu} $, with $ \tilde{\bm{R}}^0_{\bm{k},\bm{m}}=a^{-1} (\bm{r}^0_{\bm{k}}-\bm{r}^0_{\bm{m}}) $.
The symbols $ \langle \bm{k},\bm{m} \rangle$ and $ \langle\langle \bm{k},\bm{m} \rangle\rangle $ denote the sum over NNs and NNNs, respectively, while $ \langle\bm{k},\bm{p},\bm{q} \rangle$ implies that the sum is restricted to sites satisfying $ |\tilde{\bm{R}}^{0}_{\bm{k},\bm{p}}|=|\tilde{\bm{R}}^{0}_{\bm{k},\bm{q}}|=1 $. Note that, due to the presence of the factors $ \tilde{R}^0_{\bm{k},\bm{m};\mu}$, the terms $ \propto V_{3,\mu} $ strongly depend on the lattice geometry and, as we will show for the case of bipartite lattices, they give rise to anisotropic contributions in atom-atom interactions even if original dipolar forces are isotropic. 

The strength of the phonon-mediated interactions can be tailored by tuning the trapping frequencies $ \omega_\mu $ [see Eq.~\eqref{eq:V3}], which are typically of the order of hundreds kHz~\cite{Lienhard:2018,Labuhn:2016,Marcuzzi:2017,Bernien:2017}. In particular, Eq.~\eqref{eq:H2BNNN} implies that it is possible to make the overall two-body term vanish and maximize the effects of three-body interactions. Recalling Eq.~\eqref{eq:H_CS}, in order to decouple the electronic and vibration degrees of freedom and to focus only on the spin dynamics we have to require $ |W_{\bm{k},\bm{m}; \mu }| \ll \hbar \omega_\mu $. On the other hand, to access regimes governed by the effective two- and three-body interactions, one should also consider $ V_3=\sum_{\mu}V_{3,\mu} \sim V_{1,2} $. From Eqs.~\eqref{eq:W} and \eqref{eq:V3}, the above conditions translate into the following bounds on $ \omega_\mu $,
\begin{equation}\label{eq:approx_omega}
\sqrt[3]{\frac{18 \hbar}{m a^2}\left(\frac{{C_1}}{\hbar a^6}\right)^2}\ll \omega_\mu \sim \sqrt{\frac{72}{m a^2 V_{1,2}} \left(\frac{C_1}{2a^6}\right)^2}.
\end{equation}
In Fig.~\ref{fig:approx}(b), we show typical values of the effective interaction strength $ V_3 $ for a square lattice geometry. The gray region denotes the regime where $ |W_{\bm{k},\bm{m}; \mu }| \leq \hbar \omega_\mu $, while along the red dashed curve $ V_3=V_2 $. As discussed in more details in SM~\cite{Note1}, the leftmost condition in Eq.~\eqref{eq:approx_omega}, which holds for any value of $ \Omega $, can be relaxed in the strong (effective) interaction regime, where $ V_3/\Omega\gg|W_{\bm{k},\bm{m}; \mu }|/(\hbar \omega_\mu) $, while the spin-phonon decoupling becomes exact in the classical limit (i.e., with vanishing Rabi frequency $ \Omega $). Thus, as can be seen in Fig.~\ref{fig:approx}(b), the regime with $ V_2\sim V_3 $ can be  accessed experimentally and corresponds to trapping frequencies and coupling strengths achievable in Rydberg atom tweezer arrays~\cite{Marcuzzi:2017,Bernien:2017,Levine:2018}. Finally, we note that the time scales associated with the effective interaction dynamics, $ \tau_3=\hbar/V_3 $, are $ < 50\ \mu $s  in a wide region of the parameter space [i.e., the non-hatched area in Fig.~\ref{fig:approx}(b)] and are thus significantly shorter than the lifetime of the Rydberg states used in tailoring the MW-dressed potential of Fig.~\ref{fig:approx}(a), which are of the order of hundreds of $ \mu $s~\cite{Note1}.

\bigskip 

\textit{\textbf{Phase diagram for a bipartite lattice.} ---}
\begin{figure}[t]
	\centering
	\includegraphics[width=\columnwidth]{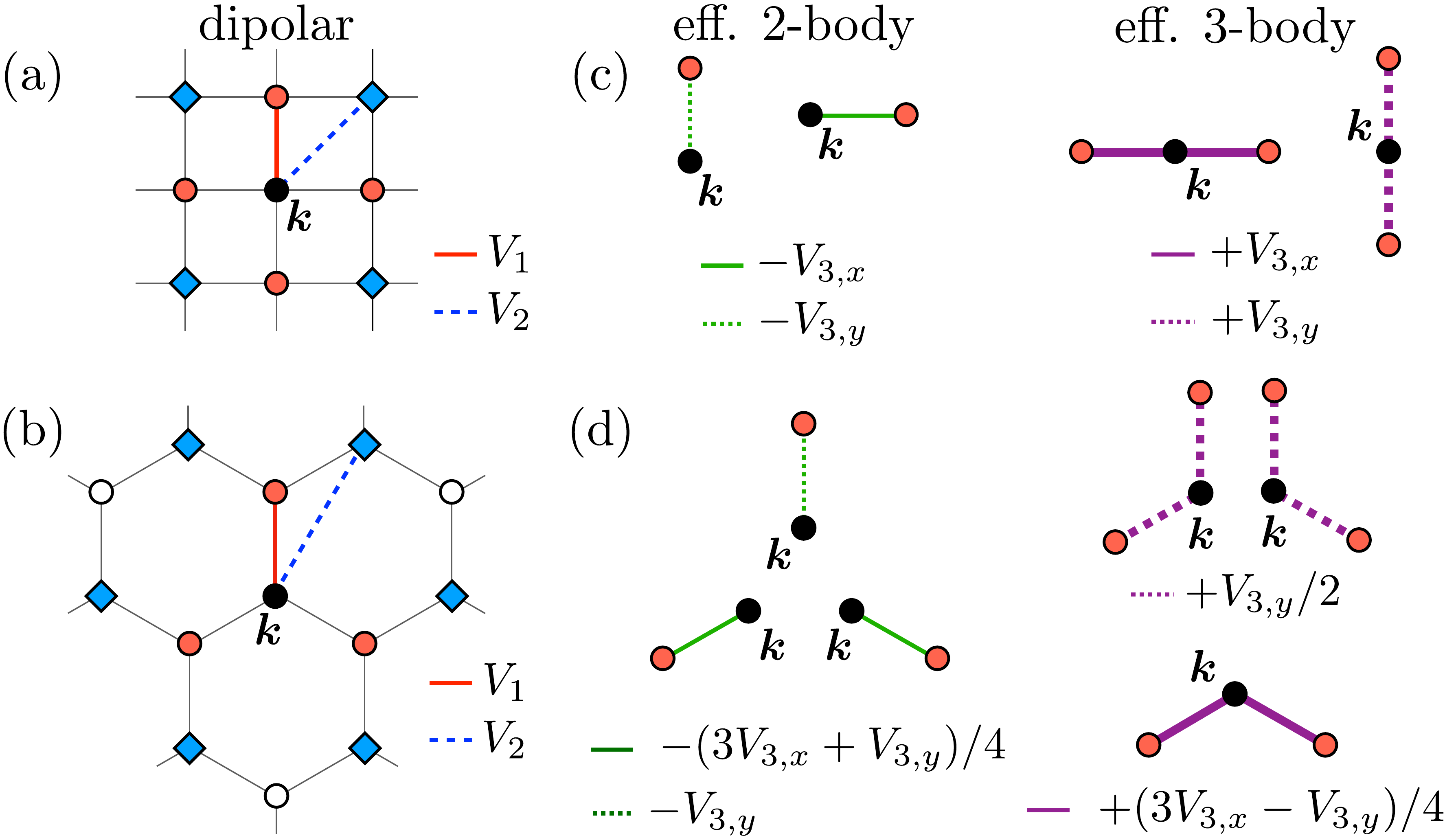}
	\caption{{\bf Interaction terms in bipartite lattices.} 
	(a) Square and (b) honeycomb lattice with NNs (orange dots) and NNNs (blue squares) interacting through dipolar interactions (red, solid and blue, dashed lines, respectively). As a consequence of the phonon-mediated effective multi-body interaction terms arise.  Two-body (green, left) and three-body (purple, right) contributions along the horizontal (solid) and vertical (dotted) direction are shown, with their corresponding sign, in panels (c) and (d) for the square and honeycomb geometries, respectively. Note that, in the latter case, horizontal (solid) terms contribute to both $ x $ and $ y $ directions, resulting in anisotropic interactions.}
	\label{fig:plaquettes}
\end{figure}
We now focus on a system of atoms arranged on a bipartite lattice and investigate the effects of the interplay between (two-body) dipolar and effective (two- and three-body) interactions on its phase diagram. The simplest case of a square lattice is shown in Fig.~\ref{fig:plaquettes}(a). The different contributions to atom-atom interaction are listed in Fig.~\ref{fig:plaquettes}(a,c). Importantly, the lattice-dependent structure of $ H_{\mathrm{3B}} $ in Eq.~\eqref{eq:H3B} implies that effective two-body interactions are attractive while, on the contrary, three-body terms have a repulsive character. This feature is quite general and, e.g., in Ising spin models on non-bipartite lattices (triangular, kagome) it could be employed to implement frustrated interactions~\cite{Balents:2010,Glaetzle:2014,Glaetzle:2015,Whitlock:2017}. The study of such phenomena will constitute the focus of future investigations. Due to the competition between two- and three-body interactions, we expect that different phases emerge. To map out the phase diagram we consider the classical limit (i.e., with vanishing Rabi frequency $ \Omega $) and determine its ground state through a classical Metropolis algorithm~\cite{Newman:1999,Binder:2010} by employing an annealing scheme~\cite{Gould:2007}. 
 \begin{figure}[t]
	\centering
	\includegraphics[width=\columnwidth]{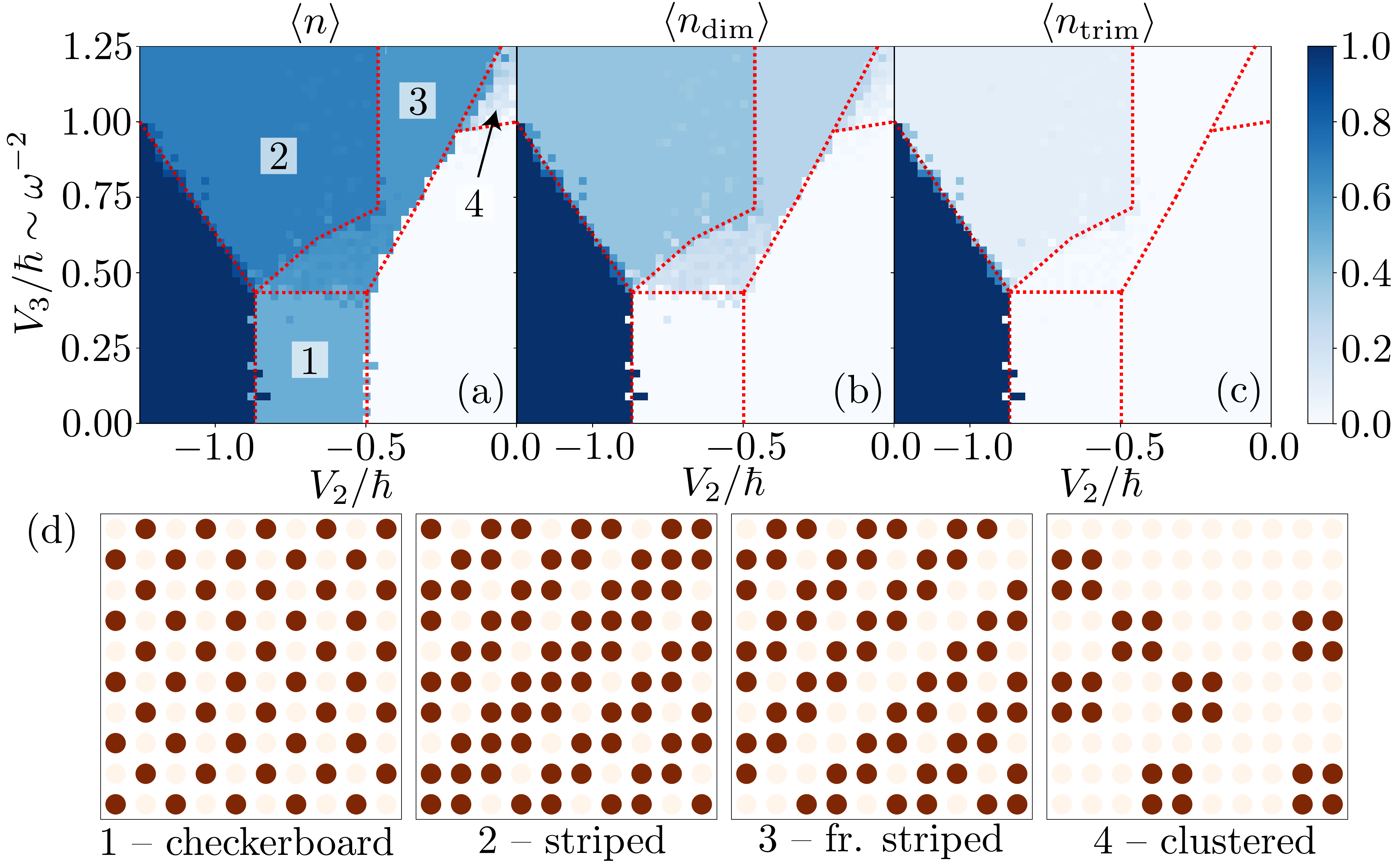}
	\caption{{\bf Phase diagram (square lattice) as a function of $ V_2/\hbar $ and $ V_3/\hbar $.} In panel (a) we show the average density of Rydberg excitations $ \av{n} $ as a function of $ V_2/\hbar $ and $ V_3/\hbar$ (units $ 2\pi\ \times $ MHz), with $V_{3,x}=V_{3,y} $, for a square lattice. Note that $ V_{3,\mu}\sim\omega_\mu^{-2} $ and, therefore, it can be controlled by the confinement strength. In panels (b) and (c) are reported the density of dimers $ \av{n_{\mathrm{dim}}} $ and trimers $ \av{n_{\mathrm{trim}}} $, respectively. Dashed red lines represent guides for the eye to distinguish between the different system phases. In panel (d) we show typical configurations in the different regions of the phase diagram. Dark (red) spots correspond to excited atoms. See text for details. In all panels, $ L=10 $, $ V_1/\hbar = 2\pi \times 0.2$ MHz and $ \Delta/\hbar= 2\pi \times 1 $ MHz. }
	\label{fig:PD_V2a_vs_V3}
\end{figure}

Results are displayed in Fig.~\ref{fig:PD_V2a_vs_V3}(a,b,c). Here, we show the behavior in the $ V_2 - V_3 $ plane (with $ V_1>0 $ and $ V_{3,x}=V_{3,y} $) of the average value of the Rydberg excitation density, $ \av{n} $, of the density of dimers $ \av{n_\mathrm{dim}} $, and of the density of trimers $ \av{n_\mathrm{trim}}  $ [see Fig.~\ref{fig:plaquettes}(c,d)]. Beyond the trivial states with all excited and all de-excited atoms, four further phases emerge, see Fig.~\ref{fig:PD_V2a_vs_V3}(d), which are: (1) checkerboard phase, dominated by the repulsive contribution $ \propto V_1 $, (2) striped phase with a single three-atom stripe, dominated by NNN two-body (attractive) interaction $ \propto V_2 $, (3) frustrated striped phase with one missing line [here, the trimers occurring in (2) are melted due to the three-body repulsive contribution $ \propto V_{3} $], and (4) four-excitation clustered phase, dominated by attractive two-body interactions $ \propto V_{3} $. Concerning this latter, we note that the transition is not as sharp as the other ones. Indeed, as can be seen from the last panel of Fig.~\ref{fig:PD_V2a_vs_V3}(d), the lattice is not entirely covered by four-particle clusters. This may suggest either that (4) is a liquid phase or that it represents a critical region. A full covering can be obtained for $ V_2>0 $, where attractive NNN interactions contribute to enhance the energy gain in forming clusters. 

Interestingly, effective interactions due to spin-phonon coupling give rise to finite-size frustration phenomena even in a square lattice in the presence of isotropic dipolar interactions. This is manifest in the emergence of the different striped phases (2) and (3): see Fig.~\ref{fig:PD_V2a_vs_V3}, which displays the case of a lattice with an even number of sites. On the contrary, if an odd number of sites is considered only a single regular striped phase emerges in this region of the phase diagram. However, a frustrated phase forms inside phase (1) (see SM~\cite{Note1}). 

In non-square lattices, the geometrical factors characterizing phonon-mediated interactions [see Eq.~\eqref{eq:Hterms_NNN}] give rise to anisotropic two- and three-body contributions even if the original dipolar interactions between atoms are isotropic. This can be seen in Fig.~\ref{fig:plaquettes}(d), where the various interaction contributions arising in a honeycomb lattice are displayed. Here, though the phase diagram is similar to the one shown in Fig.~\ref{fig:PD_V2a_vs_V3}, non-trivial and anisotropic system configurations emerge~\cite{Note1}. 

The various phases shown in Fig.~\ref{fig:PD_V2a_vs_V3} can be probed in state-of-the-art Rydberg simulators consisting of 2D defect-free arrays of optical tweezers~\cite{Barredo:2016}. Indeed, as shown in SM~\cite{Note1}, a significant part of the phase diagram in Fig.~\ref{fig:PD_V2a_vs_V3} can be mapped out by employing trapping frequencies $ \omega_\mu $ ranging from a few tens to a few hundreds kHz, while the required dipolar interaction couplings are of the order of few MHz. The desired many-body states can be prepared by real-time control of Rabi frequency and detuning via a generalization of the rapid adiabatic passage protocol proposed in Refs.~\cite{Schachenmayer:2010,Pohl:2010} and demonstrated in Ref.~\cite{Schauss:2015}. The latter is perfectly compatible with the time scales associated with the effective interactions and, in turn, with the lifetime of the Rydberg states we considered.

\bigskip 

\textit{\textbf{Conclusions.} ---} 
We have shown that electron-phonon interactions in Rydberg lattice quantum simulators permit the engineering of tunable multi-body interactions. We have illustrated the underlying mechanism in bipartite lattices, discussing in particular the case of an isotropic square lattice, where we studied the phase diagram in the classical limit. Going beyond this limit and considering the impact of quantum fluctuations ($\Omega>0$) will be possible in Rydberg quantum simulator experiments.
Many future directions of this work can be envisioned: In particular, we expect that, as a consequence of the lattice-dependent structure of the induced interactions, peculiar two- and three-body terms would arise in non-bipartite lattices (e.g., triangular, kagome), allowing for the investigation of frustrated magnetism in spin models with non-trivial multi-body interactions. Furthermore, the mechanism leading from the spin-phonon coupling to effective many-body interactions can be generalized to different kinds of bare atom-atom potentials (e.g., exchange interactions, oscillating potentials) and may allow for engineering effective interactions with different structure and/or even $ n- $body (with $ n>3 $) contributions.
 
\begin{acknowledgments}	
The research leading to these results has received funding from the European Research Council under the European Union's Seventh Framework Programme (FP/2007-2013)/ERC Grant Agreement No.~335266 (“ESCQUMA”), the EPSRC Grant No. EP/M014266/1, the EPSRC Grant No. EP/R04340X/1 via the QuantERA project “ERyQSenS”, and the Deutsche Forschungsgemeinschaft (DFG) within the SPP 1929 Giant interactions in Rydberg Systems (GiRyd). The simulations used resources provided by the University of Nottingham High-Performance Computing Service.
\end{acknowledgments}	

\footnotetext[1]{See Supplemental Material, which includes Ref.~\cite{Beterov:2009,Johansson:2012,Johansson:2013,Gambetta:2019}.}

\bibliography{3b_bibliography.bib}

\pagebreak

\widetext
\begin{center}
	\textbf{\large Supplemental Material for "Engineering non-binary Rydberg interactions via phonons in an optical lattice"}
\end{center}

\setcounter{equation}{0}
\setcounter{figure}{0}
\setcounter{table}{0}
\makeatletter
\renewcommand{\theequation}{S\arabic{equation}}
\renewcommand{\thefigure}{S\arabic{figure}}

\renewcommand{\bibnumfmt}[1]{[S#1]}

\par
\begingroup
\leftskip2cm
\rightskip\leftskip
\small In this Supplemental Material we provide additional details on the microwave dressing scheme employed to tailor the atom-atom potential in the main text. We then discuss in more details the approximations made in the main text, namely the small displacement expansion leading to Eq.~(4) and the spin-phonon decoupling in system dynamics, and we investigate their validity. We comment about the values of $ \omega_\mu $ corresponding to the phase diagram in Fig. 4 of the main text. We then discuss finite-size frustration phenomena in the phase diagram of a system arranged on a square lattice with an odd number of lattice sites. Finally, we inspect anisotropic effects in the case of a honeycomb lattice, showing its phase diagram and some typical configurations. 
\par
\endgroup

\section{Microwave dressing of Rydberg atoms}\label{SM:sec:MW}

To engineer an atom-atom interaction potential leading to dominant phonon-mediated interactions, we consider two Rydberg atoms coupled by a microwave (MW) field~\cite{Sevincli:2014,Marcuzzi:2015}. The two relevant levels on which we focus are denoted by $|r\rangle$ and $|p\rangle$. The Hamiltonian is 
\begin{equation}
	H= \sum_{j=1,2}\left[\Omega_{\mathrm{MW}} \sigma_x^{(j)}+\Delta_{\mathrm{MW}} n_p^{(j)} \right]+ \frac{C_r}{r^6}n_r^{(1)}n_r^{(2)}+\frac{C_p}{r^6}n_p^{(1)}n_{p}^{(2)} + \frac{C_3}{r^3}\left(n_r^{(1)}n_p^{(2)}+n_p^{(1)}n_r^{(2)}\right),
\end{equation}
where $\Omega_{\mathrm{MW}}$ is the MW Rabi frequency and $\Delta_{\mathrm{MW}}$ is the detuning. Here, $C_r$, $C_p$ and $C_3$ are the respective dispersion coefficients of two atoms in $|r\rangle$ and $|p\rangle$ state, and of the dipolar interaction. $n_{\sigma}^{(j)}=|\sigma^{(j)}\rangle \langle \sigma^{(j)}|$ is the density operator associated to the state $ |\sigma^{(j)}\rangle $.
\begin{figure}
	\includegraphics[width=\columnwidth]{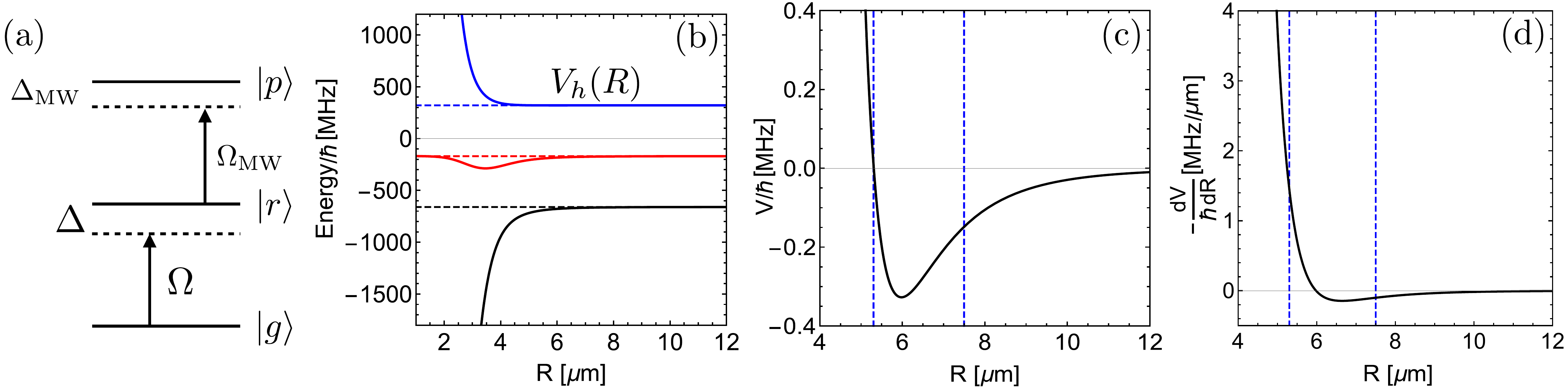}
	\caption{{\bf MW dressed potential.} (a) Each atom is modeled as a three level system, with ground state $|g\rangle$ and Rydberg states $|r\rangle$ and $|p\rangle$. The ground state $ |g\rangle $ and the Rydberg state $|r\rangle$ are laser coupled with Rabi frequency $\Omega$ and detuning $\Delta$. A MW field couples the two Rydberg states $|r\rangle$ and $|p\rangle$ with Rabi frequency $\Omega_{\mathrm{MW}}$ and detuning $\Delta_{\mathrm{MW}}$ . (b) Interaction potentials (units $ 2\pi\ \times $MHz) as a function of the inter-atom distance $ R $ (units $ \mu $m) for a pair of $ {}^{87} $Rb Rydberg atoms with $|r\rangle=|65S\rangle$, $|p\rangle = |75P\rangle$, $\Omega_{\mathrm{MW}}/\hbar = 2\pi\times230$ MHz, and $\Delta_{\mathrm{MW}}/\hbar=-2\pi\times170$ MHz. Dashed lines denote the interaction potential corresponding to $\Omega_{\mathrm{MW}}=0$. Here, we have Rydberg states $|r\rangle$ (black) and $|p\rangle$ (blue), and dipolar interaction (red). After turning on the MW coupling, the potentials are modified by the mixing of different states (solid lines). We will focus on the repulsive potential at the top of the panel,  $V_h(R)$ (blue curve). (c) Plot of the highest potential $V_h(R)$ in (a) in the presence of MW coupling. With respect to (b), the potential is shifted by $V/\hbar=2\pi\times320.4$ MHz in order to make it asymptotically vanishing. For a two-dimensional square lattice with $ a=5.3\ \mu $m, values at NNs and NNNs are $V_1=V_h(a)=2\pi\times0.01$ MHz and $V_2=V_h(\sqrt{2}a)=-2\pi \times 0.15 $ MHz, respectively. (d) The negative slope (force) of the potential shown in (c). The slope is $-2\pi\times1.45$ MHz$/\mu \mathrm{m}$ at $R=a$ and $ 2\pi\times0.10$ MHz $/\mu \mathrm{m}$ at $R=\sqrt{2}a$.}
	\label{SM:fig:potential} 
\end{figure}

By considering different Rydberg states and MW parameters it is possible to tailor the strength of the nearest neighbor (NN) and next-nearest neighbor (NNN) interactions, denoted by $ V_1 $ and $ V_2 $ respectively, and to control the slopes of the potential (forces) at these points. As stated in the main text [see Fig.~2(a) and Eq.~(10)], we are interested in a case with small values of $ V_1>0 $ and $ V_2<0 $ but with a large magnitude of the gradient of the interaction potential at NN distances, denoted by $ a $. As an example, we consider two $ {}^{87} $Rb atoms and focus on the MW dressing of Rydberg levels $|r\rangle=|65S\rangle$ and $|p\rangle = |75P\rangle$ [see Fig.~\ref{SM:fig:potential}(a)]. Their lifetime is $325.03\ \mu \mathrm{s}$ in $|r\rangle$ and $917.87\ \mu \mathrm{s}$ in $|p\rangle$ state~\cite{Beterov:2009}. The corresponding interaction strengths are $C_r /\hbar= 2\pi \times 0.37 \times 10^{3}$ GHz $\mu \mathrm{m}^6$, $C_p/\hbar=-2\pi \times 1.84 \times 10^{3}$ GHz $\mu \mathrm{m}^6$, and $C_3/\hbar=2\pi\times1.52$ MHz $\mu \mathrm{m}^3$. For a MW Rabi frequency $\Omega_{\mathrm{MW}}/\hbar = 2\pi\times230$ MHz and detuning $\Delta_{\mathrm{MW}}/\hbar=-2\pi\times170$ MHz, the higher energy potential $ V_h(R) $ at the top of panel (b), which is also shown in panel (c) and whose derivative is reported in panel (d), meets the above criteria if we consider a lattice spacing $a\approx 5.30\mu \mathrm{m}$. In particular, note that the slope at NNN distance, $ \sqrt{2}a $, is an order of magnitude smaller than the slope at $ a $. Therefore, contributions to the phonon-mediated interactions (which are proportional to the gradient of the interaction potential; see main text) arising from distances $ R>a $ can be neglected.

\section{Details on the approximations and validity checks}
We provide here additional details about the derivation of Eq.~(3) and the polaron transformation employed in the main text to decouple the spins and phonon dynamics in Eq.~(5). In particular, in the strong confinement regime, the Rydberg interaction potential $ V(\bm{r}_k,\bm{r}_m) $ can be expanded in a Taylor series to the first order in $ \delta \bm{r} $ as $ V(\bm{r}_{\bm{k}},\bm{r}_{\bm{m}})\approx V(\bm{r}^0_{\bm{k}},\bm{r}^0_{\bm{m}}) + \delta V(\bm{r}_{\bm{k}},\bm{r}_{\bm{m}})$, with
\begin{equation}\label{SM:eq:V_expansion}
	\delta V(\bm{r}_{\bm{k}},\bm{r}_{\bm{m}})=\nabla_{\bm{r}_{\bm{k}}} V(\bm{r}_{\bm{k}},\bm{r}^0_{\bm{m}})|_{\bm{r}_{\bm{k}}=\bm{r}^0_{\bm{k}}}\cdot \delta \bm{r}_{\bm{k}} +  \nabla_{\bm{r}_{\bm{m}}} V(\bm{r}_{\bm{k}}^0,\bm{r}_{\bm{m}})|_{\bm{r}_{\bm{m}}=\bm{r}^0_{\bm{m}}}\cdot \delta \bm{r}_{\bm{m}}.
\end{equation}
By substituting Eq.~\eqref{SM:eq:V_expansion} into the interaction Hamiltonian $ H_\mathrm{int} $ [Eq.~(2) of the main text] we obtain Eq.~(3) of the main text,
\begin{equation}\label{SM:eq:Hint_W}
	H_{\mathrm{int}}=\sideset{}{'}\sum_{\bm{k}, \bm{m}} \left[V^0_{\bm{k},\bm{m}}+\sum_{\mu}W_{\bm{k},\bm{m};\mu}\left(b^\dagger_{\bm{k}; \mu}+b_{\bm{k}; \mu}\right)\right] n_{\bm{k}} n_{\bm{m}},
\end{equation}
with $  V^0_{\bm{k},\bm{m}}\equiv V(\bm{r}^0_{\bm{k}},\bm{r}^0_{\bm{m}}) $, $ W_{\bm{k},\bm{m};\mu} = 2 \ell_{\mu} \left[\nabla_{\bm{r}_{\bm{k}}} V(\bm{r}_{\bm{k}},\bm{r}^0_{\bm{m}})|_{\bm{r}_{\bm{k}}=\bm{r}^0_{\bm{k}}}\right]_\mu $, and the prime in the sum meaning $ \bm{k}\neq\bm{m} $.

Since in Eq.~\eqref{SM:eq:Hint_W} the coupling between spin and phonon degrees of freedom is linear in the phonon operator $ b_{\bm{k};\mu} $, their dynamics can be decoupled by applying  to the full system Hamiltonian $ H=H_\mathrm{sp}+H_\mathrm{int} $ the canonical (polaron) transformation defined by~\cite{Porras:2004,Deng:2005}
\begin{equation}\label{SM:eq:canonical}
	U=e^{-S} \quad\text{with}\quad S=-\sum_{\bm{k},\mu} \beta_{\bm{k}; \mu} (b^\dagger_{\bm{k};\mu}-b_{\bm{k}; \mu}) \quad \text{and} \quad \beta_{\bm{k}; \mu}=n_{\bm{k}}\sum_{\bm{m}\neq\bm{k}}(\hbar \omega_\mu)^{-1} W_{\bm{k},\bm{m}; \mu }  n_{\bm{m}}.
\end{equation}
We obtain
\begin{equation}\label{SM:eq:H_CS}
	UHU^\dagger =H_\mathrm{ph}+ \underset{H_{\mathrm{spin}}}{\underbrace{\sum_{\bm{k}}\left[\Omega\sigma^x_{\bm{k}}+\Delta n_{\bm{k}}\right]+H_{\mathrm{2B}}+H_{\mathrm{3B}}}}+H_\mathrm{res}+O(\ell_{\mu}^2/a^2),
\end{equation}
with $ H_\mathrm{spin} $ the Hamiltonian governing the (decoupled) spin dynamics, $ H_\mathrm{ph}=\sum_{\bm{k},\mu} \hbar \omega_\mu b^{\dagger}_{\bm{k};\mu} b_{\bm{k};\mu} $ the phonon Hamiltonian, and $H_\mathrm{2B}$ and $H_\mathrm{3B}$ defined in Eq.~(6) of the main text. The last term is a residual spin-phonon coupling and is given by
\begin{equation}\label{SM:eq:residual}
	H_\mathrm{res}=\Omega\left(U\sum_{\bm{k}}\sigma^x_{\bm{k}}U^\dagger -\sum_{\bm{k}}\sigma^x_{\bm{k}}\right)=i\Omega\sideset{}{'}\sum_{\bm{k},\bm{m}} \sum_{\mu} \frac{W_{\bm{k},\bm{m}; \mu }}{\hbar \omega_\mu}(b^\dagger_{\bm{k};\mu}-b_{\bm{k};\mu})(\sigma^y_{\bm{k}}n_{\bm{m}}-n_{\bm{k}}\sigma^y_{\bm{m}}).
\end{equation}
Effects of this residual coupling are therefore negligible if $ |W_{\bm{k},\bm{m}; \mu }|\ll\hbar \omega_\mu $. Furthermore, two additional interesting regimes can be identified. In the classical limit $ \Omega\rightarrow 0 $ (considered in the last section of the main text), the canonical transformation results in a complete spin-phonon decoupling,
\begin{equation}\label{SM:eq:H_CS_classical}
	U H_\mathrm{classical} U^\dagger = \Delta \sum_{\bm{k}}n_{\bm{k}}+H_\mathrm{ph} +H_\mathrm{2B}+H_\mathrm{3B}+O(\ell_{\mu}^2/a^2). 
\end{equation}
On the other hand, we note that, in the strong effective two- and three-body interactions regime, the error introduced by $ H_\mathrm{res} $ can be neglected if $ V_{3,\mu}/\Omega\gg |W_{\bm{k},\bm{m}; \mu }|/\hbar \omega_\mu $. The latter condition leads to
\begin{equation}\label{key}
	\omega_\mu\ll \frac{18\hbar}{ma^2\Omega^2}\left(\frac{C_1}{a^6}\right)^2.
\end{equation}

We briefly comment here on the situation in which the trapping frequencies corresponding to the ground state, $ \omega_\mu $, and to the Rydberg state, $ \omega^\prime_\mu=\omega_\mu+\delta_\mu $ are different. Here, $ \delta_\mu $ is the trapping frequency mismatch. In this case, the phonon part of the Hamiltonian can be written as~\cite{Gambetta:2019} $ H^\prime_\mathrm{ph}=\sum_{\bm{k},\mu} \hbar \omega_\mu(1+\delta_\mu n_{\bm{k}}) b^{\dagger}_{\bm{k};\mu} b_{\bm{k};\mu} $ and, thus, an extra spin-phonon coupling contribution appears. Nevertheless, a renormalized version of the canonical transformation in Eq.~\eqref{SM:eq:canonical}, with $\beta^\prime_{\bm{k}; \mu}=(1+\delta_\mu)^{-1}n_{\bm{k}}\sum^\prime_{\bm{m}}(\hbar \omega_\mu)^{-1} W_{\bm{k},\bm{m}; \mu }  n_{\bm{m}}  $, results in a transformed Hamiltonian in which the spin terms contained in $ H_{\mathrm{spin}} $ have exactly the same structure as in Eq.~\eqref{SM:eq:H_CS}, though with renormalized coefficients. On the other hand, the residual spin-phonon coupling contribution $ H_\mathrm{res} $ of Eq.~\eqref{SM:eq:residual} now contains the extra term $ \sum_{\bm{k},\mu} \hbar \omega_\mu \delta_\mu n_{\bm{k}} b^{\dagger}_{\bm{k};\mu} b_{\bm{k};\mu} $. The latter limits the validity of our approximation up to times $ t \ll (\omega_\mu \delta_\mu)^{-1} $, which for standard experimental parameters are larger than the typical interaction induced time scales [see Fig.~\ref{SM:fig:PDomegavalues}(b)]. Moreover, ponderomotive bottle beam traps have recently been proposed for the realization of Rydberg tweezer arrays satisfying the ground-Rydberg “magic” condition~\cite{Barredo:2019}, allowing to minimize the trapping frequency mismatch $ \delta_\mu $. 

\begin{figure}
	\centering
	\includegraphics[width=\columnwidth]{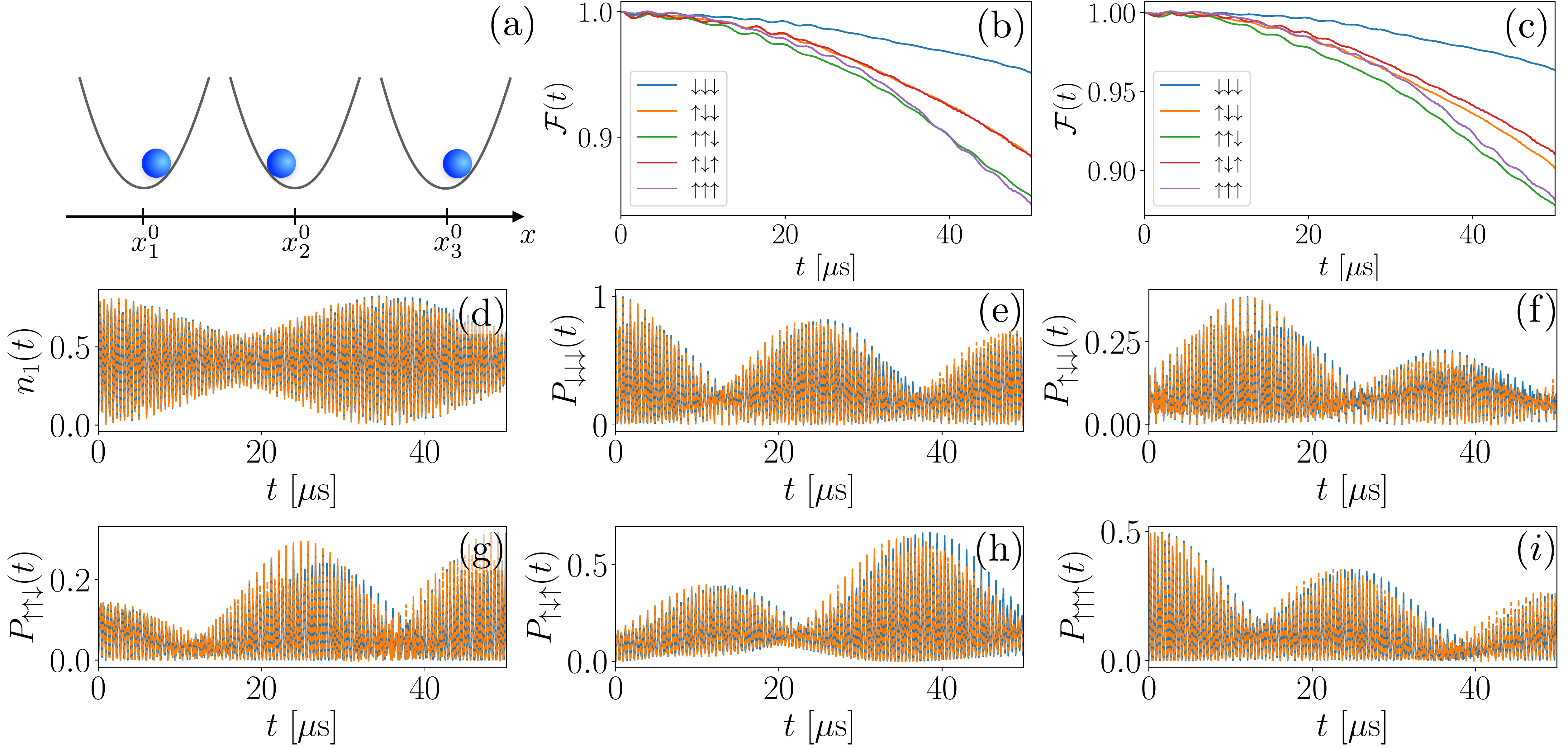}
	\caption{{\bf Full vs approximated dynamics.} (a) One dimensional three-atom setup used to verify the approximations made in the main text. For simplicity, we assume one longitudinal phonon mode per atom. (b,c) Fidelity $ \mathcal{F}(t) $, quantifying the overlap between $ \ket{\psi_{\mathrm{approx}}(t)} $ and the exact density matrix $ (\rho_\mathrm{spin}\otimes\rho_\mathrm{ph})(t) $ for the system shown in panel (a). Different curves correspond to different initial atomic configurations (see legend) while phonon modes are initialized in a thermal state with average phonon occupation (b) $ \overline{n}_{\mathrm{ph}}=0.5 $ and (c) $ \overline{n}_{\mathrm{ph}}=0.05 $. Parameters are: $ C_1/(\hbar a^6)^{-1}=2\pi\times2.5  $ MHz, $ C_2/(\hbar a_\mathrm{NNN}^6)^{-1}=-0.1C_1/(\hbar a^6)^{-1}$, $ \omega=2\pi\times 0.3 $ MHz, $ \Omega/\hbar=2\pi\times 1 $ MHz, $ \Delta/\hbar= 2\pi\times 1 $ MHz, and $ a=5 \mu $m. (d) Time evolution of the Rydberg population of atom 1 and (e-i) of the projectors $ P_{\sigma^z_1 \sigma^z_2 \sigma^z_3 }(t) $, with the phonon modes initialized in a thermal state with average phonon occupation number $ \overline{n}_{\mathrm{ph}}=0.05 $ and the spins in the $ \downarrow \downarrow \downarrow $ configuration. Blue, solid lines correspond to the approximate dynamics governed by $ H_\mathrm{spin} $, while the yellow dashed curves to the one generated by the full Hamiltonian $ H $.  Same parameters as in panels (b,c). }
	\label{SM:fig:approx}
\end{figure}

To check the validity of the approximations we made in the main text (i.e, the small displacement expansion in Eq.~(3) and the decoupling between phonon and spin dynamics in mapping out the phase diagram), we consider a minimal system of three atoms aligned along the $ x- $axis with an intermediate value of $ \Omega $, corresponding the worse case scenario. See Fig.~\ref{SM:fig:approx}(a). For the sake of simplicity, we only consider one longitudinal phonon mode per atom. We focus on the spin dynamics starting from a product state in which the phonons are initialized in a thermal state at inverse temperature $ \beta $, $ \rho_\mathrm{ph}= e^{-\beta H_\mathrm{ph}}/\mathrm{Tr}\left[e^{-\beta H_\mathrm{ph}}\right]$, and with the spins prepared in the pure state $ \rho_\mathrm{spin}=\ketbra{\psi_0}{\psi_0} $, with $ \ket{\psi_0}=\ket{s^z_1, s^z_2, s^z_3} $ and $ s^{z}_i $ the projection of the $ i- $th spin on the $ z- $axis. We then evaluate numerically~\cite{Johansson:2012,Johansson:2013} the fidelity $ \mathcal{F}(t) $, defined as the overlap between the spin state at time $ t $, $ \ket{\psi_\mathrm{approx}(t)} $, obtained in the small displacement and spin-phonon decoupling limit (i.e., with time-evolution governed by $ H_{\mathrm{spin}}=H_L+H_{\mathrm{2B}}+H_{\mathrm{3B}} $) and the full system density matrix, evolved with the full Hamiltonian $ H $ of Eq.~(2) in the main text: 
\begin{equation}\label{SM:eq:fidelity}
	\mathcal{F}(t)=\bra{\psi_{\mathrm{approx}}(t)} e^{-iHt} (\rho_\mathrm{spin} \otimes \rho_\mathrm{ph}) e^{iHt} \ket{\psi_{\mathrm{approx}}(t)}.
\end{equation}

\begin{figure}
	\centering
	\includegraphics[width=\columnwidth]{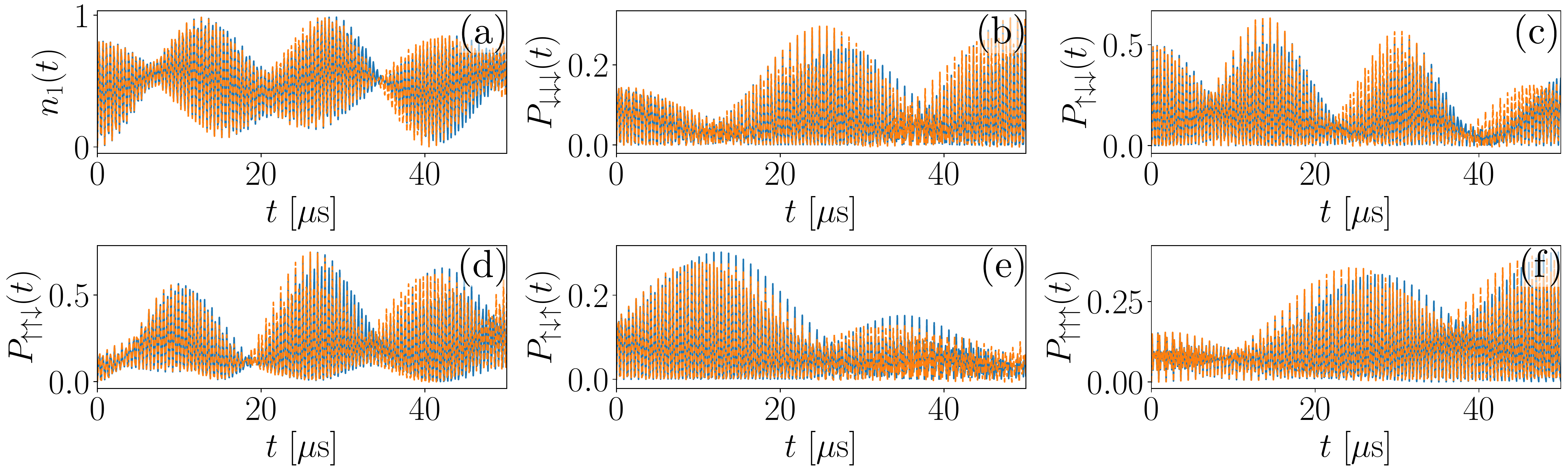}
	\caption{{\bf Full vs approximated dynamics - spin initialized in an excited state}. (a-f) Same as Fig~\ref{SM:fig:approx}(d-i) with spins initialized in the $ \downarrow\uparrow\uparrow $ configuration.}
	\label{SM:fig:approx2}
\end{figure}

As can be seen from Fig.~\ref{SM:fig:approx}(b,c), $ \mathcal{F}(t) $ remains close to $ 1 $ over a significant time window, especially when small temperatures are considered [panel (b)]. Here, the latter have been chosen in order to correspond to an average phonon occupation number $ \overline{n}_{\mathrm{ph}}=0.5 $ [panel (b)] and $ \overline{n}_{\mathrm{ph}}=0.05 $ [panel (c)], which can be achieved in optical tweezer setups via Raman sideband cooling  ~\cite{Kaufman:2012,Thompson:2013}. 
In Fig.~\ref{SM:fig:approx}(d-i) and Fig.~\ref{SM:fig:approx2} we compare the time evolution of the expectation values of the population of the Rydberg state of atom 1, $ n_1 $, and of various projectors on states with $ \sigma^z_1$, $ \sigma^z_2$, $ \sigma^z_3 $ spin components along the $ z- $direction, e.g., $ P_{\uparrow \uparrow \uparrow}=\av{n_1 n_2 n_3} $, for the dynamics governed by $ H $ and $ H_\mathrm{spin} $ for two different initial spin configurations. In both cases, the agreement between the approximate and full dynamics is excellent. 

\section{Trapping frequencies range, effective interactions time scales and classical ground state preparation}
\begin{figure}
	\centering
	\includegraphics[width=0.8\columnwidth]{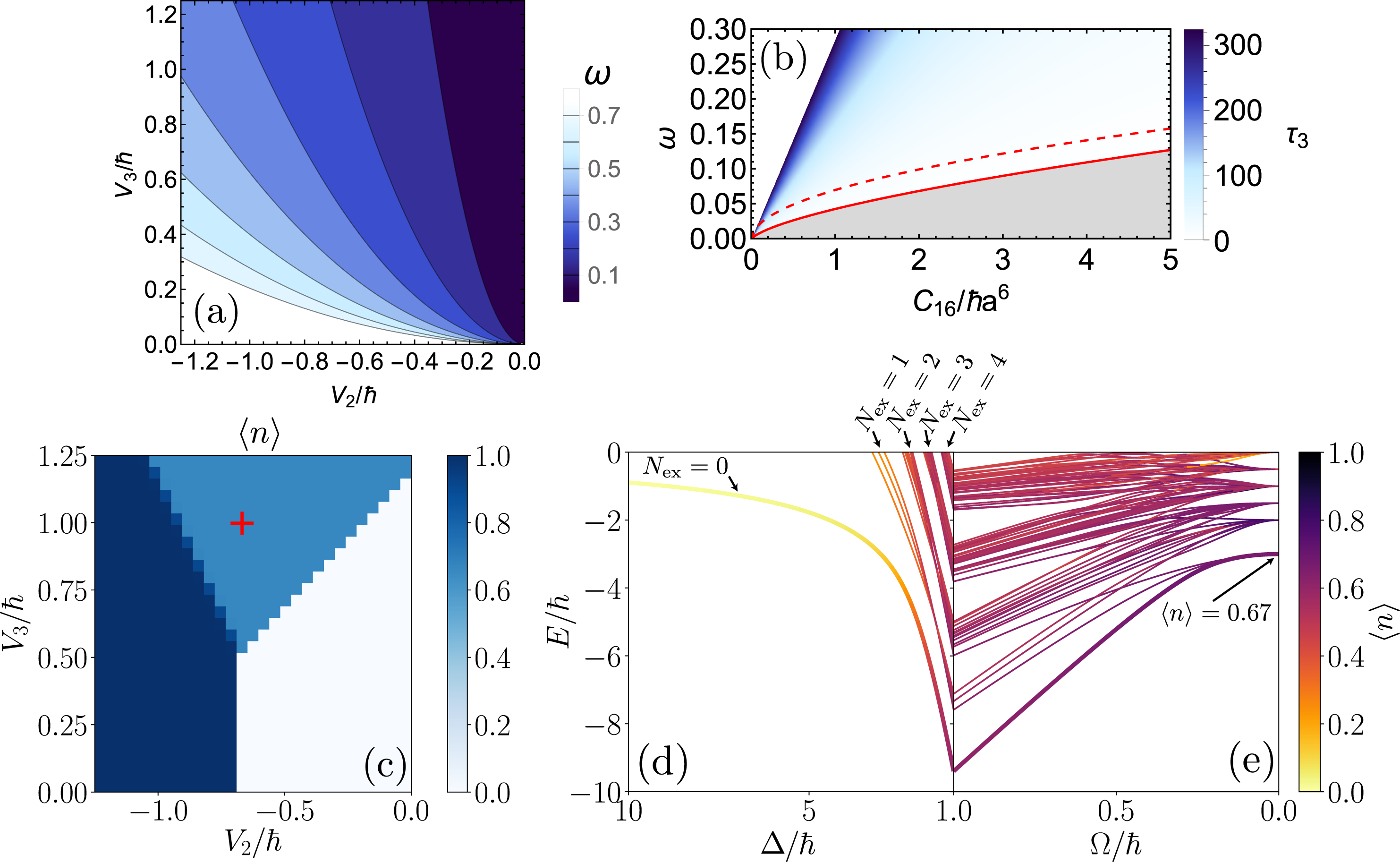}
	\caption{{\bf Frequency range, effective interaction time scales, and ground state preparation.} (a) Density plot in the $ V_2-V_3 $ plane of the trapping frequencies (units $ 2\pi\ \times $ MHz) corresponding to the values of the effective interaction $ V_3/\hbar $ on the $ y- $axis. (b) Density plot of the time scale $ \tau_3=\hbar/V_3 $ associated with the effective interaction $ V_3/\hbar $ in the $ C_1/(\hbar a^6) - \omega $ plane (units $ 2\pi\ \times $ MHz). As in Fig.~2(b) in main text, the gray area denotes the regime in which $ |W_{\bm{k},\bm{m}; \mu }| \leq \hbar \omega_\mu $ (with equality corresponding to the red solid line). Along the red dashed curve $ V_3=V_{2} $. Here, $ C_2/(\hbar a_\mathrm{NNN}^6)=-0.1 C_1/(\hbar a^6) $. (c) Exact phase diagram for a $ 3\times3 $ square lattice spin system for $ V_1/\hbar = 2\pi \times 0.2$ MHz, $ V_2/\hbar = -2\pi \times 0.7$ MHz, $ V_3/\hbar = 2\pi \times 1$ MHz, $ \Omega/\hbar= 2\pi \times 0.01 $ MHz, and $ \Delta/\hbar= 2\pi \times 1 $ MHz [corresponding to the red cross in panel (c)]. (d, e) Many-body energy spectrum of $ H_{\mathrm{spin}} $ for a $ 3\times3 $ square lattice as a function of (d) $ \Delta/\hbar $ and (e) $ \Omega/\hbar $ (units $ 2\pi\ \times $ MHz). Here, $ \Delta_{\mathrm{in}}/\hbar=2\pi \times 10 $ MHz, $ \Delta_{\mathrm{fin}}/\hbar=2\pi \times 1 $ MHz, $ \Omega_{\mathrm{in}}/\hbar=2\pi \times 1 $ MHz, and $ \Omega_{\mathrm{fin}}=0 $.The colormap indicates the average excitation density $ \av{n} $ in the various eigenstates. The lowest, thick line denotes the ground state. For a large value of the detuning $ \Delta $ the spectrum separates in different manifolds with fixed number of excited atoms (see labels). In the limit of vanishing $ \Omega $ the ground state becomes $ 6- $fold degenerate, where each one of the $ 6 $ eigenstates corresponds to a specific arrangement of the $ N_{\mathrm{ex}}=5 $ excited atoms on the $ 3\times3 $ lattice. }
	\label{SM:fig:PDomegavalues}
\end{figure}
In Fig.~\ref{SM:fig:PDomegavalues}(a) we reproduce the $ V_2 - V_3 $ plane of Fig.~4 in the main text and show the values of the trapping frequencies $ \omega_\mu $ corresponding to the values of $ V_3 $ used in drawing the system phase diagram. As can be seen, in the majority of the $ V_2-V_3 $ plane, the values of $ \omega_\mu $ are both within the approximation validity regimes shown in Fig.~2(a) and within experimentally achievable values. Figure~\ref{SM:fig:PDomegavalues}(b) displays the time scale associated with the effective two- and three-body interactions, i.e. $ 1/\tau_3=V_3/\hbar $, which are considerably lower than the lifetime of the Rydberg levels considered in engineering the MW dressed potential in Section~\ref{SM:sec:MW}. 

To illustrate how the classical ground states of the system shown in Fig.~4 of the main text can be prepared, we focus on a $ 3\times3 $ spin square lattice described by $ H_\mathrm{spin} $. Its phase diagram, obtained through exact diagonalization and corresponding to the same parameters used in Fig.~4, is shown in Fig.~\ref{SM:fig:PDomegavalues}(c). Note that, as a consequence of the reduced system size, the only non-trivial phase emerging here is the striped one, [phase (2) in Fig.~4]. As an example, the ground state corresponding to the parameters identified by the red cross in Fig.~\ref{SM:fig:PDomegavalues}(c) can be prepared by following the rapid adiabatic passage procedure described in Ref.~\cite{Schauss:2015}: By setting a large value of the detuning $ \Delta_{\mathrm{in}} $ and a vanishing value of the Rabi frequency $ \Omega $, the system is initialized at first in a state with all the atoms in their ground state (i.e., with $ \av{n}=0 $). Then, $ \Omega $ is slowly increased up to $ \Omega_{\mathrm{in}} $, inducing a coupling between states belonging to the various manifolds with different Rydberg excitation density. Subsequently, the detuning is decreased from $ \Delta_{\mathrm{in}} $ to $ \Delta_{\mathrm{fin}} $ and, finally, $ \Omega $ is reduced  to its final value $ \Omega_{\mathrm{fin}}=0 $. As can be seen from the lowest, thick curve in panels (d) and (e), this procedure leads to a degenerate ground state with the desired value of $ \av{n} $ and well-seperated from high-lying many-body eigenstates. Here, the ground state degeneracy is associated with all the possible arrangements of the excited atoms. As shown in Ref.~\cite{Schauss:2015}, an overall laser sweep duration of a few $ \mu $s is sufficient to ensure a good degree of adiabaticity and, therefore, this protocol is perfectly compatible with the typical time scales of our system [see, e.g., Fig.~\ref{SM:fig:PDomegavalues}(b)].

\section{Phase diagram for a square lattice with odd number of sites}
\begin{figure}
	\centering
	\includegraphics[width=0.7\columnwidth]{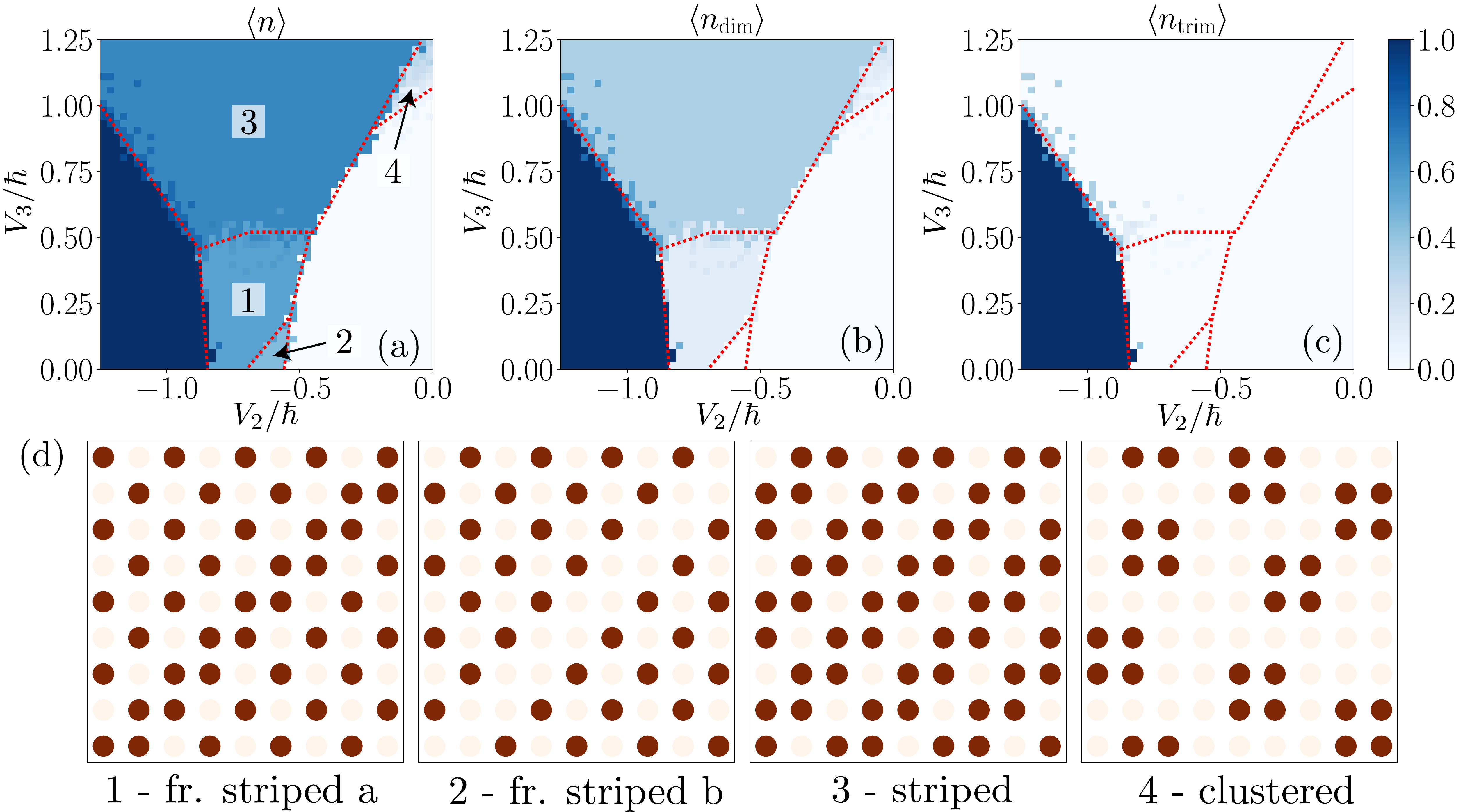}
	\caption{{\bf Phase diagram for a square lattice with an odd number of lattice sites as a function of $ V_2 $ and $ V_3 $.} In panel (a) we show the average density of Rydberg excitations $ \av{n} $ as a function of $ V_2/\hbar $ and $ V_{3,x}/\hbar=V_{3,y}/\hbar $ (units $ 2\pi\ \times $ MHz) for a square lattice. In panels (b) and (c) are reported the density of dimers $ \av{n_{\mathrm{dim}}} $ and trimers $ \av{n_{\mathrm{trim}}} $ [see Fig.~3(b) in main text for details], respectively. Dashed red lines represent guides for the eye to distinguish between the different system phases. In panel (d) we show typical configurations in the different regions of the phase diagram. Dark (red) spots correspond to excited atoms. In all panels, $ L=9 $, $ V_1/\hbar = 2\pi \times0.2$ MHz and $ \Delta/\hbar=2\pi \times 1 $ MHz. }
	\label{SM:fig:SL_odd}
\end{figure}

In the main text we have investigated the phase diagram of a system of atoms arranged on a square lattice with an even number of sites (see Fig.~4 in the main text). In that case, different phases are present and, in particular, finite-size frustration phenomena emerge [see, e.g., phases (2) and (3) in Fig. 4 in the main text]. Here, we inspect what happens if a square lattice with an odd number of sites is considered. The phase diagram and some typical system configurations are shown in Fig.~\ref{SM:fig:SL_odd}. The two (frustrated) phases (2) and (3) of the even number of sites case are now merged [see phase (3) in Fig.~\ref{SM:fig:SL_odd}(a)] in a single phase showing regular strips of dimers. However, finite-size frustration phenomena emerge in the small $ V_3 $ region of the phase diagram: see phases (1) and (2) in Fig.~\ref{SM:fig:SL_odd}, which consists of stripes of single excitations with a (a) dimer or (b) a missing strip, respectively. We note that the four-excitation clustered phase occurs in this case as well and exhibits a behavior similar to the one discussed in the main text. 

\section{Phase diagram for a honeycomb lattice}
\begin{figure}
	\centering
	\includegraphics[width=0.9\columnwidth]{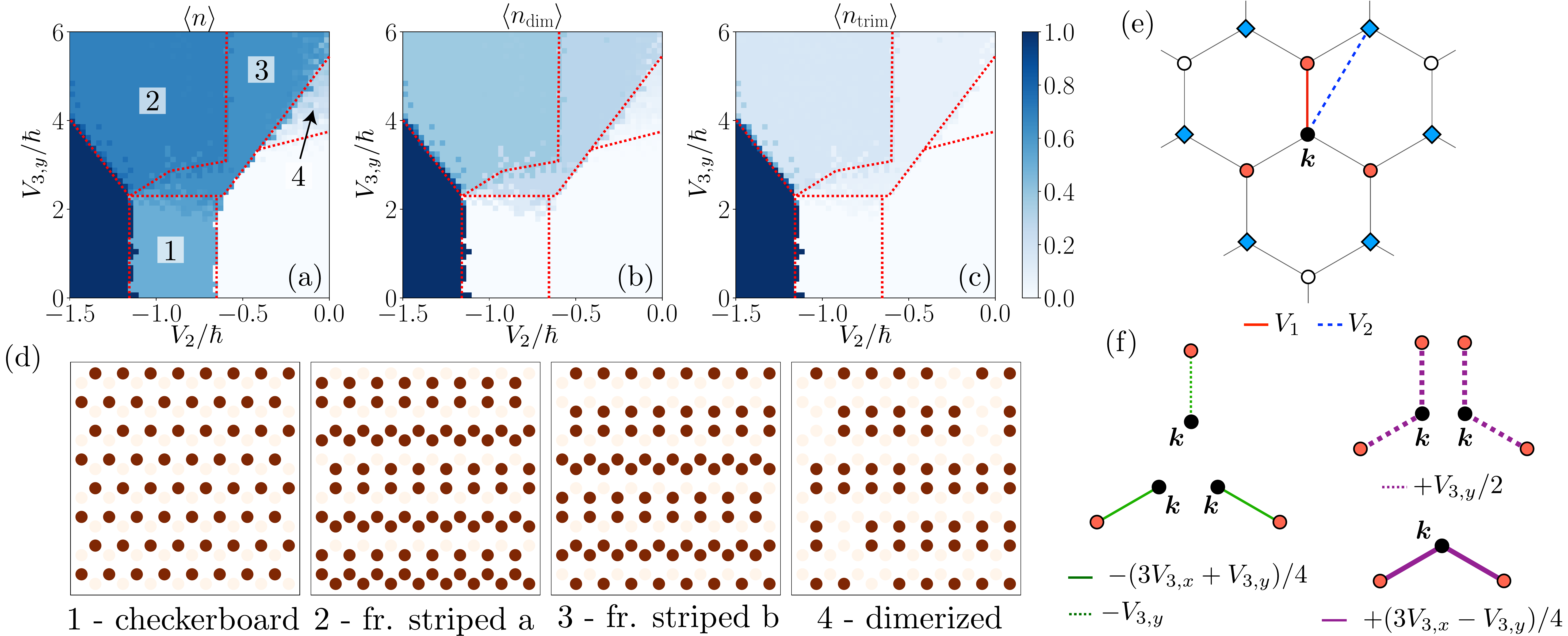}
	\caption{{\bf Phase diagram for a honeycomb lattice as a function of $ V_2 $ and $ V_3 $.} In panel (a) we show the average density of Rydberg excitations $ \av{n} $ as a function of $ V_2/\hbar $ and $ V_{3,y}/\hbar $ (units $ 2\pi\ \times$ MHz) with $ V_{3,x}/\hbar= 2\pi\ \times1 $ MHz for a honeycomb lattice with $ L^2=64 $ unit cells (and $ 2L^2=128 $ sites). In panels (b) and (c) we display the density of dimers $ \av{n_{\mathrm{dim}}} $ and trimers $ \av{n_{\mathrm{trim}}} $ [as defined in panel (f)], respectively. Dashed red lines represent guides for the eye to distinguish between the different system phases. In panel (d) we show typical configurations in the different regions of the phase diagram. Dark (red) spots correspond to excited atoms. In panels (a-d), $ L=8 $, $ V_1/\hbar = 2\pi \times0.5$ MHz and $ \Delta/\hbar=2\pi \times 2 $ MHz. (e) Honeycomb lattice with NNs (orange dots) and NNNs (blue squares) interacting through dipolar interactions (red, solid and blue, dashed lines, respectively). (f) Two-body (green, left) and three-body (purple, right) phonon-mediated contributions along the horizontal (solid) and vertical (dotted) direction. }
	\label{SM:fig:HC}
\end{figure}

In this section we finally investigate the phase diagram for a honeycomb lattice. In this case, as stated in the main text, the geometrical factors present in Eq. (11) can be exploited to realize non-trivial and anisotropic interactions [see Fig.~3(b) of the main text and Fig.~\ref{SM:fig:HC}(e,f)]. As an example, Figs.~\ref{SM:fig:HC}(a,b,c) show the phase diagram in the $ V_2 - V_{3,y}$ plane (with $ V_1>0 $) of a system with $ L^2=64 $ unit cells (and $ 2L^2=128 $ sites)  with fixed $ V_{3,x}/\hbar=2\pi\times 1 $ MHz. Even if the phase diagram looks similar to the one obtained for a square lattice and discussed in Fig.~4 of the main text, anisotropic phenomena can be seen in typical system configurations [see Fig.~\ref{SM:fig:HC}(d)]. In particular, as can be seen from Fig.~\ref{SM:fig:HC}(f), with the previous choice of parameters (e.g., with $ V_{3,x}/\hbar=2\pi\times 1 $ MHz) the three-body interaction along the $ x- $direction changes from repulsive to attractive when $ V_{3,y}=3V_{3,x} $.  Above this threshold we thus expect the formation of “chains” of excitations along $ x $ to be energetically preferred. Indeed, the emergence of such structures is clearly visible in the typical configurations associated to phases (2) and (3) in Fig.~\ref{SM:fig:HC}(d). 

\end{document}